# CHARACTERISTIC SOLUTIONS OF THE CHAIN OF VLASOV EQUATIONS

**E.E. Perepelkin[a,c,d,*], B.I. Sadovnikov[a], N.G. Inozemtseva[b,c], A.S. Medvedev[a]**

[a]*Faculty of Physics, Lomonosov Moscow State University, Moscow, 119991 Russia*
[b]*Moscow Technical University of Communications and Informatics, Moscow, 123423 Russia*
[c]*Dubna State University, Moscow region, Dubna, 141980 Russia*
[d]*Joint Institute for Nuclear Research, Moscow region, Dubna,141980 Russia*
**Corresponding author: pevgeny@jinr.ru*

**Abstract**

A new method has been presented of constructing a class of exact solutions of an infinite self-linking chain of the Vlasov equations for distribution functions of kinematic quantities of all orders. Using the characteristic transformation of variables proposed in this paper, any equation from the Vlasov chain can be reduced to the mathematical form of the first Vlasov equation. Since the solution of the first Vlasov equation can be found by the solution of the Schrödinger equation, the authors have proposed an algorithm for constructing characteristic solutions for an arbitrary equation from the Vlasov chain.

The proposed method of construction of exact solutions has been successfully implemented on an example of time-dependent quantum system with thermodynamic parameter in the form of inverse temperature. These found exact solutions are also applicable to quantum dot systems.

## Introduction

A central concept in physics is the equation of motion, which can be a consequence of some conservation law or symmetry principle [1-4]. The equation of motion $\Lambda\left(\vec{r},\vec{v},\dot{\vec{v}},\ddot{\vec{v}},...\right)=0$ contains a different set of kinematic quantities $\vec{r},\vec{v},\dot{\vec{v}},\ddot{\vec{v}},...$ coordinate, velocity, acceleration, second-order acceleration, and so on. For example, one of the first known equations of motion belongs to Aristotle (4th century BC) $m\kappa\vec{v}-\vec{F}=0$, where $m$ is mass, $\vec{F}$ is force and $\kappa$ is an empirical coefficient. Increasing accuracy in measuring technology led to Newton's second law (1687) $m\dot{\vec{v}}-\vec{F}=0$. The study of electromagnetism yielded Maxwell's equations and Larmor's formula for the power of radiation of an electromagnetic wave by a particle with charge $q$, moving with acceleration. This gave rise to the Lorentz-Abraham-Dirac equation of motion (1892) $m\tau_0\ddot{\vec{v}}-m\dot{\vec{v}}+\vec{F}=0$, where $\tau_0=q^2/6\pi\varepsilon_0 mc^3$. In 1860 M.V. Ostrogradsky proposed mechanics in which the Euler-Lagrange equation is written for the Lagrangian function $L\left(\vec{r},\vec{v},\dot{\vec{v}},\ddot{\vec{v}},...,\vec{v}^{(n-1)},t\right)$ of $n$ kinematic quantities.

Analyzing the history of the evolution of the equations of motion, we can note several important points. First, the order of differential equations of motion has been constantly increasing. Aristotle (1st order), Newton (2nd order), Lorentz-Abraham-Dirac (3rd order) and Ostrogradsky ($n$ - order). Second, the equations of motion were a consequence of experimental observations. As the accuracy of measurements improved or new properties of matter (charge, electromagnetic field) were discovered, the equations of motion were modified. The phenomenological nature of the equations of motion is due to the fact that we do not fully know all the laws of nature. For example, to date, the area of cosmology related to dark matter and dark energy requires special consideration [5-8]. Third, is it certain that there is a finite order of the equation and what should it be? Note that from a mathematical point of view, small parameters at higher derivatives in a differential equation can introduce substantial instability (uncertainty) into the behavior of the solution. There are many examples here, ranging from dynamical systems theory [9-11] to quantum mechanics [12, 13].

On the other hand, the question of the kinematic trajectory being a solution of the equation of motion can be reformulated in a different form. In the middle of the twentieth century, A.A. Vlasov considered [14, 15] an infinite-dimensional generalized phase space (GPS) containing the complete set of kinematic quantities of all orders $\Omega_\infty = \left\{\vec{r}, \vec{v}, \dot{\vec{v}}, \ddot{\vec{v}}, ...\right\}$. For the point $\vec{\xi}_0 = \left(\vec{r}_0, \vec{v}_0, \dot{\vec{v}}_0, \ddot{\vec{v}}_0, ...\right)^T \in \Omega_\infty$ one can formally (under the condition of convergence) construct the Taylor series

$$\vec{\xi}(t) = e^{t\hat{D}}\vec{\xi}_0 = M_\infty(t)\vec{\xi}_0, \tag{i.1}$$

$$M_{N_a}(t) \overset{\text{def}}{=} \begin{pmatrix} 1 & t & t^2/2 & t^3/3!.. \\ 0 & 1 & t & t^2/2... \\ 0 & 0 & 1 & t... \\ ... & ... & ... & .... \\ 0 & ... & ... & 0 \;\; 1 \end{pmatrix}, \quad M_\infty = \lim_{N_a \to +\infty} M_{N_a}, \quad \det M_\infty = 1,$$

where $\vec{u}_\xi \overset{\text{def}}{=} \hat{D}\vec{\xi} \overset{\text{def}}{=} \left\{\vec{v}, \dot{\vec{v}}, \ddot{\vec{v}}, ...\right\}^T$ is the generalized velocity vector field. In fact, the representation (i.1) defines a one-parameter Lie group. A unique generalized phase trajectory (i.1) passes through each point of $\vec{\xi}_0 \in \Omega_\infty$ GPS. Due to the uniqueness of the Taylor series expansion, the trajectories (i.1) do not intersect in the GPS. The GPS contains all analytic trajectories of motion possible in nature. A.A. Vlasov defined the distribution function $f_\infty\left(\vec{\xi}_0, t\right)$ satisfying the first principle the law of conservation of probabilities, in GSP

$$\frac{\partial f_\infty}{\partial t} + \operatorname{div}_\xi \left( f_\infty \vec{u}_\xi \right) = 0, \tag{i.2}$$

where $\operatorname{div}_\xi = \operatorname{div}_r + \operatorname{div}_v + \operatorname{div}_{\dot{v}} + ...$ Continual integration of equation (i.2) over phase subspaces gives a self-linking chain of Vlasov equations, which can be written in a compact form [16]

$$\hat{\pi}_n S_n = -Q_n, \; n \in \mathbb{N}, \tag{i.3}$$

$$\hat{\pi}_n \overset{\text{def}}{=} \frac{\partial}{\partial t} + \vec{v} \cdot \nabla_r + \dot{\vec{v}} \cdot \nabla_v + ... + \left\langle \vec{\xi}^{n+1} \right\rangle \cdot \nabla_{\xi^n}, \quad S_n \overset{\text{def}}{=} \operatorname{Ln} f_n, \quad Q_n \overset{\text{def}}{=} \operatorname{div}_{\xi^n} \left\langle \vec{\xi}^{n+1} \right\rangle, \tag{i.4}$$

$$f_n\left(\vec{\xi}^{1,...,n}, t\right) \overset{\text{def}}{=} \int_{\mathbb{R}^3} ... \int_{\mathbb{R}^3} f_\infty\left(\vec{\xi}, t\right) \prod_{s=n+1}^{+\infty} d^3\xi^s, \qquad \left\langle \vec{\xi}^{n+1} \right\rangle \overset{\text{def}}{=} \frac{1}{f_n} \int_{\mathbb{R}^3} f_{n+1}\left(\vec{\xi}^{1,..,n+1}, t\right) \vec{\xi}^{n+1} d^3\xi^{n+1}, \tag{i.5}$$

where $\vec{\xi}^{1,...,n} \overset{\text{def}}{=} \left(\vec{r}, \vec{v}, ...\vec{v}^{(n-2)}\right)^T$. The value $f_0$ can be interpreted as the total number of particles in the system or as a normalization factor for the probability density function. The operator $\hat{\pi}_n$ (i.4) defines the total time derivative along the generalized phase trajectory (i.1). With $n = 0$ it is implied that $\hat{\pi}_0 = d/dt$. The value $Q_n$ (i.4) defines the dissipation source density. As a result, it follows from equation (i.3) that the change in the distribution density $f_n$ along the phase trajectory is equal to the density of dissipation sources $-Q_n$. If there are no dissipation sources ($Q_n = 0$) then $\hat{\pi}_n f_n = 0$ and $f_n = const$ along the phase trajectory. Thus, at $n=2$ and $m\left\langle \dot{\vec{v}} \right\rangle = -\nabla_r U$ one obtains the well-known Liouville equation for the phase space $\Omega_2 = \{\vec{r}, \vec{v}\}$. In the general case at $n=2$ the second Vlasov equation is known in plasma physics, statistical physics, astrophysics, and accelerator physics [17-23].

Note that the original equation (i.2) has no dissipation sources since $Q_\infty = \operatorname{div}_\xi \vec{u}_\xi = 0$. The result $\operatorname{div}_\xi \vec{u}_\xi = 0$ follows from the uniqueness of the Taylor series expansion (i.1), i.e. there are no points $\vec{\xi} \in \Omega_\infty$ from which many trajectories exit (enter). When averaging (i.5) some of the kinematic information about the system described by the distribution function $f_\infty$ is lost since there is a transition to the function $f_n$. Averaging over higher kinematic quantities (i.5) leads to the possibility of existence of dissipation sources $Q_n$ (i.4).

The chain of Vlasov equations (i.3) is connected with the chain of equations for $H_n$ - Boltzmann functions

$$\hat{\pi}_0 [f_0 H_n] = -f_0 \left\langle \ldots \left\langle Q_n \right\rangle \ldots \right\rangle, \; n \in \mathbb{N}, \tag{i.6}$$

$$H_n(t) \stackrel{\text{def}}{=} \frac{1}{f_0} \int_{\mathbb{R}^3} \ldots \int_{\mathbb{R}^3} f_n\left(\vec{\xi}^{1,\ldots,n}, t\right) S_n \prod_{s=1}^{n} d^3\xi^s = \left\langle \ldots \left\langle S_n \right\rangle \ldots \right\rangle (t). \tag{i.7}$$

In the special case for the classical phase space ($n=2$) the function (i.7) corresponds to the well-known $H$ - Boltzmann function. It follows from expression (i.6) that the evolution of $H_n$ - Boltzmann functions is determined by the average sources of dissipations.

Another fundamental property of the Vlasov chain (i.3) is the existence of three types of conservation laws, which can be conventionally called the laws of conservation of «mass», «momentum» and «energy».

The law of conservation of «mass» is obtained by integrating $n$ -th equation of the chain by $\int \ldots d^3\xi^n$. As a result, the equation for the function $f_n$ is followed by the equation for the function $f_{n-1}$. Such a sequence of integrations can reach $f_0$ defining the total mass / charge / number of particles or normalizing the total probability to one.

The law of conservation of «momentum» is obtained for each equation in the chain (i.3) by multiplying it by the kinematic quantity component $\xi_\lambda^n$ and then integrating over $\int \ldots d^3\xi^n$ [16]

$$\hat{\pi}_n \left\langle \xi_k^{n+1} \right\rangle = \left[ \frac{\partial}{\partial t} + \left\langle \xi_s^{n+1} \right\rangle \frac{\partial}{\partial \xi_s^n} \right] \left\langle \xi_k^{n+1} \right\rangle = -\frac{1}{f_n} \frac{\partial P_{ks}^{n+1}}{\partial \xi_s^n} + \left\langle \left\langle \xi_k^{n+2} \right\rangle \right\rangle, \tag{i.8}$$

$$P_{ks}^{n+1} \stackrel{\text{def}}{=} \int_{\mathbb{R}^3} \left( \xi_k^{n+1} - \left\langle \xi_k^{n+1} \right\rangle \right) \left( \xi_s^{n+1} - \left\langle \xi_s^{n+1} \right\rangle \right) f_{n+1} d^3\xi^{n+1}, \tag{i.9}$$

where the quantity (i.9) conventionally corresponds to the pressure tensor in the hydrodynamic approximation. Hence, $\frac{1}{f_n} \frac{\partial P_{ks}^{n+1}}{\partial \xi_s^n}$ is the conventional pressure force and $\left\langle \left\langle \xi_k^{n+2} \right\rangle \right\rangle$ is conventionally responsible for the external force acting on the system. The expression on the left (i.8) denotes the time derivative of the kinematic quantity. For example, in the special case at $n=1$ the law of momentum conservation (i.8) will take the following form

$$\hat{\pi}_2 \left\langle v_k \right\rangle = \left[ \frac{\partial}{\partial t} + \left\langle v_s \right\rangle \frac{\partial}{\partial x_s} \right] \left\langle v_k \right\rangle = -\frac{1}{f_1} \frac{\partial P^2_{ks}}{\partial x_s} + \left\langle \left\langle \dot{v}_k \right\rangle \right\rangle. \tag{i.10}$$

The equation of motion (i.10) is well known in continuum mechanics.

The law of conservation of «energy» is obtained from each equation of the chain (i.3) by multiplying it by $\left(\xi^n\right)^2$ and then integrating by $\int ... d^3\xi^n$

$$\frac{\partial}{\partial t}\left[\frac{f_n}{2}\left\langle \xi^{n+1}\right\rangle^2 + \frac{1}{2}\mathrm{Tr}\, P^{n+1}_{kk}\right] + \frac{\partial}{\partial \xi^n_s}\left[\frac{f_n}{2}\left\langle \xi^{n+1}\right\rangle^2 \left\langle \xi^{n+1}_s\right\rangle + \frac{1}{2}\left\langle \xi^{n+1}_s\right\rangle \mathrm{Tr}\, P^{n+1}_{kk} + \left\langle \xi^{n+1}_k\right\rangle P^{n+1}_{ks} + \frac{1}{2}\mathrm{Tr}\, P^{n+1}_{kks}\right] =$$
$$= \int_{\mathbb{R}^3} f_{n+1}\left\langle \xi^{n+2}_k \right\rangle \xi^{n+1}_k \, d^3\xi^{n+1}, \tag{i.10}$$

$$P^{n+1}_{ksl} \overset{\mathrm{def}}{=} \int_{\mathbb{R}^3} \left(\xi^{n+1}_k - \left\langle \xi^{n+1}_k\right\rangle\right)\left(\xi^{n+1}_s - \left\langle \xi^{n+1}_s\right\rangle\right)\left(\xi^{n+1}_l - \left\langle \xi^{n+1}_l\right\rangle\right) f_{n+1} d^3\xi^{n+1}. \tag{i.11}$$

The first summand in the left-hand side of equation (i.10) corresponds to the time variation of the «energy» density. The second summand in the left part of (i.10) is the divergence from the current density of «energy». In the right part of equation (i.10) stands the work of external «forces». In the special case at $n=1$ , the law of conservation of energy (i.11) takes the following form

$$\frac{\partial}{\partial t}\left[\frac{f_1}{2}\left|\left\langle \vec{v}\right\rangle\right|^2 + \frac{1}{2}\mathrm{Tr}\, P^2_{kk}\right] + \frac{\partial}{\partial x_s}\left[\frac{f_1}{2}\left|\left\langle \vec{v}\right\rangle\right|^2 \left\langle v_s\right\rangle + \frac{1}{2}\left\langle v_s\right\rangle \mathrm{Tr}\, P^2_{kk} + \left\langle v_k\right\rangle P^2_{ks} + \frac{1}{2}\mathrm{Tr}\, P^2_{kks}\right] =$$
$$= \int_{\mathbb{R}^3} f_2 \left\langle \dot{v}_k \right\rangle v_k \, d^3 v. \tag{i.12}$$

The summand $\frac{f_1}{2}\left|\left\langle \vec{v}\right\rangle\right|^2$ in equation (i.12) determines the density of kinetic energy and $\frac{1}{2}\mathrm{Tr}\, P^2_{kk}$ is the density of the internal energy; $\frac{f_1}{2}\left|\left\langle \vec{v}\right\rangle\right|^2 \left\langle v_s\right\rangle$ is the current density of kinetic energy and $\frac{1}{2}\left\langle v_s\right\rangle \mathrm{Tr}\, P^2_{kk}$ is the current density of the internal energy; $\left\langle v_k\right\rangle P^2_{ks}$ characterizes the work of gravitational forces; $\frac{1}{2}\mathrm{Tr}\, P^2_{kks}$ corresponds to the heat flow [14].

The conservation laws (i.8), (i.10) are valid for each equation of the Vlasov chain, i.e. there are infinitely many of them. Each of the conservation laws is written for a certain set of kinematic quantities that describe the system. The amount of necessary kinematic information is determined by the level of accuracy in describing the physical system and taking into account its types of interaction. This is the situation in the historical retrospective given above (from Aristotle to the present day). The kinematic chain of the Vlasov equations contains complete information about the system in the form of a distribution function. The conservation laws are not phenomenological, but follow from one single first principle, the law of conservation of probabilities (i.2) in the GPS.

Note that the Vlasov chain is self-linking, i.e. finding its solution requires cut it off at some equation containing the necessary amount of kinematic information about the system. The chain is cut off by introducing a dynamic approximation of the average kinematic quantities (i.5). In [24] the Schrödinger equation, the Hamilton-Jacobi equation and the equation of motion of a charged particle in an electromagnetic field were derived by cut the chain off on the first equation. In [25], this result is extended to self-consistent systems for which the Maxwell equations and the Pauli and Dirac equation are derived. Using the Wigner-Vlasov formalism the Vlasov-Moyal approximation was constructed in [13], which allowed to cut the chain off on the second equation and reduce it to the known Moyal equation [26] for the Wigner function [27, 28] of a quantum system in phase space. In [29] a variant of the extension of the Vlasov-Moyal approximation for electromagnetic systems is considered. In [30, 31] a general approach in the construction of quantum-mechanical analogs of the Schrödinger and Hamilton-Jacobi equations for systems described by higher kinematic quantities is considered.

The Wigner-Vlasov formalism mathematically rigorously connects solutions of quantum equations with solutions of classical equations of continuum mechanics, field theory, statistical physics, thermodynamics, accelerator physics, solid state physics and plasma physics.

The purpose of this work is to find a specific transformation that allows in a particular case to reduce $n$-th equation from the chain to a simpler equation for which the exact solution is known. Knowing such a transformation it is possible to obtain exact solutions of $n$-th Vlasov chain equation by exact solutions of the simple equation.

The paper has the following structure. In §1 we construct the characteristic transformation of variables for $n$-th equation from the Vlasov chain. As a result, $n$-th equation is reduced to the mathematical form of the first Vlasov equation. The solution of the first Vlasov equation is sought by reducing it to the analogue of the Schrödinger equation for a particle in an electromagnetic field. Thus, knowing the exact solution of the Schrödinger equation it is possible to construct the characteristic solution of $n$-th Vlasov equation. In §2 we consider a quantum system with a thermodynamic parameter in the form of inverse temperature [32] for which we construct the exact solution of $n$- th Vlasov equation. The conclusion contains a brief review of the obtained results. Intermediate mathematical transformations are given in Appendix.

## §1 Characteristic transform

Let the volume of kinematic information be chosen to describe some physical system. In general case it is possible to correspond to such a system $n$-th equation from the Vlasov chain (i.3). At the same time, to cut the chain off on equations, there must exist a dynamic approximation of the mean flux of the kinematic quantity $\left\langle \vec{\xi}^{n+1} \right\rangle$ (i.5). The following theorem is valid.

**Theorem 1** *$n$ -th time-dependent equation from the infinite chain of Vlasov equations (i.3)*

$$\frac{\partial f_n}{\partial t} + \operatorname{div}_r\left[f_n \vec{v}\right] + \operatorname{div}_v\left[f_n \dot{\vec{v}}\right] + \ldots + \operatorname{div}_{\overset{(n-2)}{v}}\left[f_n \left\langle \overset{(n-1)}{\vec{v}} \right\rangle\right] = 0, \qquad (1.1)$$

*admits reduction to the analog of the first Vlasov equation for function $F_n\left(\vec{\eta}_n, \tau_n\right)$*

$$\frac{\partial F_n}{\partial \tau_n} + \operatorname{div}_{\eta_n}\left[F_n \left\langle \overset{(n-1)}{\vec{u}} \right\rangle\right] = 0, \qquad (1.2)$$

*by introducing the characteristic*

$$\vec{\eta}_n\left(\vec{\xi}^{1,\ldots,n}, t\right) \overset{\text{def}}{=} -\sum_{k=0}^{n-1} \tau_k(t) \overset{(k)}{\vec{r}}, \quad \tau_n(t) \overset{\text{def}}{=} (-1)^{n+1} \frac{t^n}{n!}, \qquad (1.3)$$

*if the representations are valid*

$$f_n\left(\vec{\xi}^{1,\dots,n},t\right)\overset{\text{def}}{=}F_n\left(\vec{\eta}_n,\tau_n\right),\quad \left\langle \overset{(n-1)}{\vec{v}}\right\rangle\left(\vec{\xi}^{1,\dots,n},t\right)\overset{\text{def}}{=}\left\langle \overset{(n-1)}{\vec{u}}\right\rangle\left(\vec{\eta}_n,\tau_n\right). \tag{1.4}$$

The proof of Theorem 1 is given in Appendix.

**Remark 1**

Note that the function $f_n\left(\vec{\xi}^{1,\dots,n},t\right)$ is defined in $3n,\ n\geq 2$ phase space $\Omega_n$ and the function $F_n\left(\vec{\eta}_n,\tau_n\right)$ in coordinate space $\mathbb{R}^3\in\vec{\eta}_n$. The variables $t$ and $\tau_n$ correspond to time, which is a parameter for phase trajectories in the space $\Omega_n$.

The solution $F_n$ of equation (1.2) may not depend on the time variable $\tau_n$ (that is $F_n=F_n\left(\vec{\eta}_n\right)$), but the solution $f_n$ of equation (1.1) will depend on time $t$. The converse is not true for the transformation (1.3). Thus, equation (1.1) is always time-dependent.

It follows from Theorem 1 that knowing the solution of equation (1.2) we can construct the characteristic solution of $n$-th Vlasov equation.

Let us represent the phase space $\Omega_n$ as a direct sum of projection spaces $\Omega_n=\Omega_n^x\times\Omega_n^y\times\Omega_n^z$. The dimensionality of each of the spaces $\Omega_n^x,\Omega_n^y,\Omega_n^z$ is equal to $n$. When $\vec{\eta}_n=\vec{\eta}_{n,0}$ is fixed expression (1.3) defines three hyperplane equations for the spaces $\Omega_n^x,\Omega_n^y,\Omega_n^z$

$$\Omega_n^{\ell}:\ \eta_{n,0}^{(\ell)}+\sum_{k=0}^{n-1}\tau_k\overset{(k)}{\ell}=0,\quad \ell=x,y,z. \tag{1.5}$$

The coefficients $\tau_n$ in equations (1.5) correspond to the normal vector to the hyperplane. Since $\tau_n=\tau_n(t)$ (1.3) the hyperplanes (1.5) will rotate with time $t$. Given fixed components of the vector $\vec{\eta}_{n,0}$ on each hyperplane (1.5) function $F_n=const$. The characteristic hyperplanes corresponding to the same time $t$ but different $\vec{\eta}_{n,0}$ are coplanar since they have the same normal vectors with components $\tau_k(t)$ (1.5).

Consider the phase domain $\mathrm{P}\subset\Omega_n$ as a multidimensional parallelepiped with some faces defined by characteristic hyperplanes (1.5) corresponding to vectors $\pm\vec{\eta}_{n,0}$. Let us calculate the following integral of the distribution function $f_n\left(\vec{\xi}^{1,\dots,n},t\right)$ over the domain $\mathrm{P}=\mathrm{P}^x\times\mathrm{P}^y\times\mathrm{P}^z$

$$\int_{\mathrm{P}} f_n\left(\vec{\xi}^{1,\dots,n},t\right)\prod_{k=0}^{n-1}d^3\overset{(k)}{r}=\int_{\mathrm{P}^z}\prod_{k=0}^{n-1}d\overset{(k)}{z}\int_{\mathrm{P}^y}\prod_{j=0}^{n-1}d\overset{(j)}{y}\int_{\mathrm{P}^x}\mathrm{F}_n\left(\eta_n^{(x)},\eta_n^{(y)},\eta_n^{(z)},\tau_n\right)\prod_{l=0}^{n-1}d\overset{(l)}{x}, \tag{1.6}$$

where $F_n\left(\vec{\eta}_n,\tau_n\right)\overset{\text{def}}{=}\mathrm{F}_n\left(\eta_n^{(x)},\eta_n^{(y)},\eta_n^{(z)},\tau_n\right)$. Let us transform the repeated integral (1.6). For the subspace $\mathrm{P}^x$, we obtain

$$\int_{P^x} F_n\left(\eta_n^{(x)},\eta_n^{(y)},\eta_n^{(z)},\tau_n\right)\prod_{l=0}^{n-1} d\overset{(l)}{x} = \int_{-\overset{(n-1)}{x_0}}^{\overset{(n-1)}{x_0}} d\overset{(n-1)}{x}\dots\int_{-\dot{x}_0}^{\dot{x}_0} d\dot{x} \int_{-x_0+\sum_{s=1}^{n-1}\tau_s \overset{(s)}{x}}^{x_0+\sum_{s=1}^{n-1}\tau_s\overset{(s)}{x}} F_n\left(x-\sum_{s=1}^{n-1}\tau_s\overset{(s)}{x},\eta_n^{(y)},\eta_n^{(z)},\tau_n\right)dx =$$

$$= \int_{-\overset{(n-1)}{x_0}}^{\overset{(n-1)}{x_0}} d\overset{(n-1)}{x}\dots\int_{-\dot{x}_0}^{\dot{x}_0} d\dot{x}\int_{-x_0}^{x_0} F_n\left(\eta_n^{(x)},\eta_n^{(y)},\eta_n^{(z)},\tau_n\right)d\eta_n^{(x)} = 2^{n-1}\prod_{l=1}^{n-1}\overset{(l)}{x_0}\int_{-x_0}^{x_0} F_n\left(\eta_n^{(x)},\eta_n^{(y)},\eta_n^{(z)},\tau_n\right)d\eta_n^{(x)}. \quad (1.7)$$

When integrating over the subspaces $P^y, P^z$ we can perform actions similar to (1.7). As a result, the integral (1.6) will take the form

$$\int_P f_n\left(\vec{\xi}^{1,..,n},t\right)\prod_{k=0}^{n-1} d^3\overset{(k)}{r} = 8^{n-1}M_n\left(\tau_n\right)\prod_{j=1}^{n-1}\overset{(j)}{x_0}\,\overset{(j)}{y_0}\,\overset{(j)}{z_0}, \quad (1.8)$$

$$M_n\left(\tau_n\right) \overset{\text{def}}{=} \int_{-z_0}^{z_0} d\eta_n^{(z)}\int_{-y_0}^{y_0} d\eta_n^{(y)}\int_{-x_0}^{x_0} F_n\left(\eta_n^{(x)},\eta_n^{(y)},\eta_n^{(z)},\tau_n\right)d\eta_n^{(x)}. \quad (1.9)$$

If $F_n$ is explicitly independent of the time parameter $\tau_n$ then the value (1.9) is constant and we can introduce a normalization condition for the time-dependent function $f_n\left(\vec{\xi}^{1,\dots,n},t\right)$ in the bounded region $P$ (1.8)

$$f_n \mapsto N_n f_n, \quad N_n^{-1} = 8^{n-1}M_n\prod_{j=1}^{n-1}\prod_{\ell}\overset{(j)}{\ell_0}. \quad (1.10)$$

As will be shown in §2 there exist solutions of $F_n$ with an explicit dependence on $\tau_n$, but their norm does not depend on $\tau_n$.

Consider the transition from the distribution function $f_n$ to the function $f_{n-1}$ (see Appendix)

$$f_{n-1}\left(\vec{\xi}^{1,\dots,n-1},t\right) = \int_{-\overset{(n-1)}{z_0}}^{\overset{(n-1)}{z_0}} d\overset{(n-1)}{z}\int_{-\overset{(n-1)}{y_0}}^{\overset{(n-1)}{y_0}} d\overset{(n-1)}{y}\int_{-\overset{(n-1)}{x_0}}^{\overset{(n-1)}{x_0}} f_n\left(\vec{\xi}^{1,..,n},t\right)d\overset{(n-1)}{x},$$

hence

$$f_{n-1}\left(\vec{\xi}^{1,\dots,n-1},t\right) = \frac{1}{\tau_{n-1}^3}\int_{\eta_{n,-}^{(z)}\left(\vec{\xi}^{1,\dots,n-1},t\right)}^{\eta_{n,+}^{(z)}\left(\vec{\xi}^{1,\dots,n-1},t\right)} d\eta_n^{(z)}\int_{\eta_{n,-}^{(y)}\left(\vec{\xi}^{1,\dots,n-1},t\right)}^{\eta_{n,+}^{(y)}\left(\vec{\xi}^{1,\dots,n-1},t\right)} d\eta_n^{(y)}\int_{\eta_{n,-}^{(x)}\left(\vec{\xi}^{1,\dots,n-1},t\right)}^{\eta_{n,+}^{(x)}\left(\vec{\xi}^{1,\dots,n-1},t\right)} F_n\left(\eta_n^{(x)},\eta_n^{(y)},\eta_n^{(z)},\tau_n\right)d\eta_n^{(x)}, \quad (1.11)$$

where

$$\vec{\eta}_{n,\pm}\left(\vec{\xi}^{1,\dots,n-1},t\right) \overset{\text{def}}{=} -\sum_{s=0}^{n-2}\tau_s\overset{(s)}{\vec{r}} \pm \tau_{n-1}(t)\overset{(n-1)}{\vec{r}_0} = \vec{\eta}_{n-1}\left(\vec{\xi}^{1,\dots,n-1},t\right) \pm \tau_{n-1}(t)\overset{(n-1)}{\vec{r}_0}. \quad (1.12)$$

The limits of integration in expression (1.11) are functions of phase coordinates (1.12) $\vec{\xi}^{n-1}$. Re-integration of expression (1.11) over the kinematic quantity space $\int d^3\xi^{n-1}$ will result in the distribution function $f_{n-2}$. The average kinematic quantities are found in the same way $\left\langle \overset{(n-2)}{\vec{v}} \right\rangle, \left\langle \overset{(n-3)}{\vec{v}} \right\rangle, \ldots$

For equation (1.2) we can construct an analog of the Schrödinger equation. Let us decompose the vector field $\left\langle \overset{(n-1)}{\vec{u}} \right\rangle$ by Helmholtz's theorem into a vortex $\vec{A}_n$ and a vortex-free component $\nabla_{\eta_n} \Phi_n$

$$\left\langle \overset{(n-1)}{\vec{u}} \right\rangle (\vec{\eta}_n, \tau_n) = -\alpha_n \nabla_{\eta_n} \Phi_n (\vec{\eta}_n, \tau_n) + \gamma_n \vec{A}_n (\vec{\eta}_n, \tau_n), \tag{1.13}$$

where $\alpha_n, \gamma_n$ are some constant values. For a positive density function $F_n(\vec{\eta}_n, \tau_n)$ we introduce the notation $F_n = |\Psi_n|^2$, where $\Psi_n(\vec{\eta}_n, \tau_n) = |\Psi_n| e^{i\varphi_n} \in \mathbb{C}$. According to [24, 25] we associate the phase $\varphi_n(\vec{\eta}_n, \tau_n)$ with the scalar potential $\Phi_n$

$$\Phi_n = 2\varphi_n + 2\pi k, \; k \in \mathbb{Z},$$
$$\left\langle \overset{(n-1)}{\vec{u}} \right\rangle = i\alpha_n \nabla_{\eta_n} \left( \ln \left| \frac{\Psi_n}{\Psi_n^*} \right| + i \,\mathrm{Arg} \frac{\Psi_n}{\Psi_n^*} \right) + \gamma_n \vec{A}_n = i\alpha_n \nabla_{\eta_n} \mathrm{Ln} \frac{\Psi_n}{\Psi_n^*} + \gamma_n \vec{A}_n. \tag{1.14}$$

We will consider the vortex field $\vec{A}_n$ to be defined with the accuracy of $\Psi$ - Lorenz gauge [25], i.e. $\mathrm{div}_{\eta_n} \vec{A}_n = g$, where $g$ is some function defined below. As a result, the following theorem is valid.

**Theorem 2** *Let the function $F_n(\vec{\eta}_n, \tau_n)$ be positive $F_n = |\Psi_n|^2$ and satisfy equation (1.2) and the mean flux $\left\langle \overset{(n-1)}{\vec{u}} \right\rangle$ admits the Helmholtz decomposition (1.14) then for $\Psi_n(\vec{\eta}_n, \tau_n) \in \mathbb{C}$ is true equation*

$$\frac{i}{\beta_n} \frac{\partial}{\partial \tau_n} \Psi_n = -\alpha_n \beta_n \left( \hat{\mathrm{p}}_n - \frac{\gamma_n}{2\alpha_n \beta_n} \vec{A}_n \right)^2 \Psi_n + U_n \Psi_n, \tag{1.15}$$

*where $\hat{\mathrm{p}}_n \overset{\mathrm{def}}{=} \frac{i}{\beta_n} \nabla_{\eta_n}$ $\beta_n$ is a constant value and $U_n(\vec{\eta}_n, \tau_n) \in \mathbb{R}$ is some function. In this cas, the analog of the Hamilton-Jacobi equation is satisfied*

$$-\frac{1}{\beta_n} \frac{\partial \varphi_n}{\partial \tau_n} = -\frac{1}{2\alpha_n \beta_n} \left| \left\langle \overset{(n-1)}{\vec{u}} \right\rangle \right|^2 + \mathrm{V}_n = \mathrm{H}_n, \tag{1.16}$$

*where*

$$\mathrm{V}_n = U_n + \mathrm{Q}_n, \; \mathrm{Q}_n = \frac{\alpha_n}{\beta_n} \frac{\Delta_{\eta_n} |\Psi_n|}{|\Psi_n|}, \tag{1.17}$$

*and equation (1.16) corresponds to the equation of motion*

$$\hat{\pi}_1 \left\langle \overset{(n-1)}{\vec{u}} \right\rangle = \left( \frac{\partial}{\partial \tau_n} + \left\langle \overset{(n-1)}{\vec{u}} \right\rangle \cdot \nabla_{\eta_n} \right) \left\langle \overset{(n-1)}{\vec{u}} \right\rangle = -\gamma_n \left( \vec{E}_n + \left\langle \overset{(n-1)}{\vec{u}} \right\rangle \times \vec{B}_n \right), \tag{1.18}$$

$$\vec{E}_n \overset{\text{def}}{=} -\frac{\partial \vec{A}_n}{\partial \tau_n} - \frac{2\alpha_n \beta_n}{\gamma_n} \nabla_{\eta_n} \mathrm{V}_n, \quad \vec{B}_n \overset{\text{def}}{=} \mathrm{curl}_{\eta_n} \vec{A}_n. \tag{1.19}$$

*The scalar* $\mathrm{V}_n$ *and vector* $\vec{A}_n$ *potentials satisfy the* $\Psi$ *- Lorenz gauge*

$$\mathrm{div}_{\eta_n} \vec{A}_n + \frac{2\alpha_n \beta_n}{\gamma_n} \frac{1}{c^2} \frac{\partial \mathrm{V}_n}{\partial \tau_n} = 0. \tag{1.20}$$

The proof of Theorem 2 coincides with the proof of the theorems from [24, 25].

**Remark 2**

From the formal point of view by choosing $\alpha_n = -\hbar_n/2m$, $\beta_n = 1/\hbar_n$, $\gamma_n = -q_n/m$ equation (1.15) will be similar to the Schrödinger equation for an electromagnetic system, equation (1.16) will change into the Hamilton-Jacobi equation and equation (1.18) describes the motion of a charged particle in the electromagnetic field (1.19). The view of the potential $\mathrm{Q}_n$ (1.17) coincides with the quantum potential in the «wave-pilot» theory [33, 34], and the gauge (1.20) decays into two gauges for the potentials (1.17) $\mathrm{V}_n = U_n + \mathrm{Q}_n$ and $\vec{A}_n = \vec{A}_n^U + \vec{A}_n^{\mathrm{Q}}$

$$\mathrm{div}_{\eta_n} \vec{A}_n^U + \frac{1}{q_n c^2} \frac{\partial U_n}{\partial \tau_n} = 0, \; \mathrm{div}_{\eta_n} \vec{A}_n^{\mathrm{Q}} + \frac{1}{q_n c^2} \frac{\partial \mathrm{Q}_n}{\partial \tau_n} = 0. \tag{1.21}$$

We note again that the above «coincidence» is mathematically formal since in [24, 25] the expressions (1.15)-(1.20) are derived from the first Vlasov equation for the coordinate wave function. In our case, the function $\Psi_n$ is related to the distribution density of the characteristic quantity $\vec{\eta}_n$ (1.3) in $3n$-dimensional phase space $\Omega_n$. Nevertheless, the similarity of mathematical equations allows us to construct characteristic solutions of the Vlasov equation (1.1) by exact solutions of the analog of the Schrödinger equation (1.15).

## §2 Exact solutions

Let us consider an example of constructing an exact characteristic solution of the Vlasov equation (1.1) on the basis of the material from §1. The general principle is as follows. For given potentials $U_n$ and $\vec{A}_n$ the solution $\Psi_n$ of equation (1.15) is found. Knowing $\Psi_n$ the function $F_n = |\Psi_n|^2$ and the mean flux (1.14) are determined. As a result, the solution of equation (1.2) is found. Using Theorem 1 by the solution of (1.2) we construct (1.4) the solution $f_n$ of equation (1.1).

We take $\Omega_n = \Omega_n^x = \mathrm{P}^x$ as the phase space. Equation (1.1) will take the form

$$\frac{\partial f_n}{\partial t} + v \frac{\partial f_n}{\partial x} + \dot{v} \frac{\partial f_n}{\partial v} + \ldots + \frac{\partial}{\partial \overset{(n-2)}{v}} \left[ f_n \left\langle \overset{(n-1)}{v} \right\rangle \right] = 0, \tag{2.1}$$

which according to Theorem 1 under condition (1.4) reduces to the form of the first Vlasov equation

$$\frac{\partial F_n}{\partial \tau_n}+\frac{\partial F_n}{\partial \eta_n}\left[F_n\left\langle \overset{(n-1)}{u}\right\rangle\right]=0, \tag{2.2}$$

for the function $F_n(\eta_n,\tau_n)$, where $\eta_n=-\sum_{k=0}^{n-1}\tau_k(t)\overset{(k)}{x}$. It follows from Theorem 2 that equation (2.2) is related to the analog of the Schrödinger equation (1.15). Due to the presence of only one projection of the phase domain $\Omega_n^x$ the vector potential $\vec{A}_n$ is excluded from consideration and we take the scalar potential $U_n$ as an infinitely deep well of width $a_n$

$$U_n(\eta_n)=\begin{cases}0, \ if \ \ 0<\eta_n<a_n,\\ +\infty, \ otherwise.\end{cases} \tag{2.3}$$

The boundary value problem for the analog of the Schrödinger equation (1.15) with potential (2.3) takes the form

$$\begin{cases} i\dfrac{\partial}{\partial \tau_n}\Psi_n=-\dfrac{\hbar_n}{2m}\dfrac{\partial^2}{\partial \eta_n^2}\Psi_n, \ 0<\eta_n<a_n,\\ \Psi_n(0,\tau_n)=\Psi_n(a_n,\tau_n)=0,\end{cases} \tag{2.4}$$

and its solution

$$\Psi_n^{\mu}(\eta_n,\tau_n)=\sqrt{\frac{2}{a_n}}\sin\left(\frac{\sqrt{2mE_\mu}}{\hbar_n}\eta_n\right)e^{-i\frac{E_\mu}{\hbar_n}\tau_n}, \ E_\mu=\frac{\pi^2\hbar_n^2\mu^2}{2ma_n^2}, \ \mu\in\mathbb{N}, \tag{2.5}$$

where $\mu$ is the number of the «quantum» state. From (2.5), (1.14) and (1.3)-(1.4) it follows that the solution of equation (2.2) has the form

$$F_n^{\mu}(\eta_n)=\frac{2}{a_n}\sin^2\left(\frac{\sqrt{2mE_\mu}}{\hbar_n}\eta_n\right), \qquad \left\langle \overset{(n-1)}{u}\right\rangle=0. \tag{2.6}$$

Let us write the normalization condition (1.7)-(1.10)

$$\int_{\mathrm{P}^x} F_n^{\mu}\prod_{l=0}^{n-1}d\overset{(l)}{x}=\int_{\overset{(n-1)}{\Delta a_n}} d\overset{(n-1)}{x}\ldots\int_{\Delta \dot{a}_n} d\dot{x}\int_{\sum_{s=1}^{n-1}\tau_s\overset{(s)}{x}}^{a_n+\sum_{s=1}^{n-1}\tau_s\overset{(s)}{x}} F_n^{\mu}\left(x-\sum_{s=1}^{n-1}\tau_s\overset{(s)}{x}\right)dx=\prod_{l=1}^{n-1}\overset{(l)}{\Delta a_n}\int_0^{a_n}F_n^{\mu}(\eta_n)d\eta_n,$$

$$N^{-1}=\prod_{l=1}^{n-1}\overset{(l)}{\Delta a_n}, \tag{2.7}$$

where $\overset{(l)}{\Delta a_n}$ are the linear dimensions of the regions in the corresponding phase subspaces. By virtue of (2.6)-(2.7) equation (2.1) and its solution will take the following form

$$\frac{\partial f_n^\mu}{\partial t}+v\frac{\partial f_n^\mu}{\partial x}+\dot{v}\frac{\partial f_n^\mu}{\partial v}+...+\overset{(n-2)}{v}\frac{\partial f_n^\mu}{\partial \overset{(n-3)}{v}}=0, \tag{2.8}$$

$$f_n^\mu\left(x,\dot{x},\ddot{x},...,\overset{(n-1)}{x},t\right)=2\left(\prod_{l=0}^{n-1}\overset{(l)}{\Delta a_n}\right)^{-1}\sin^2\left(\frac{\sqrt{2mE_\mu}}{\hbar_n}\sum_{k=0}^{n-1}\overset{(k)}{x}\frac{(-1)^k t^k}{k!}\right), \tag{2.9}$$

where $\Delta a_n \overset{\text{def}}{=} a_n$. The absence of the flow (2.6) leads to a substantial simplification of the original equation (2.1) and a transition to equation (2.8). In the case of (2.8) the solution (2.9) can be expanded

$$f_n\left(\xi^{1,...,n},t\right)=G\left(\eta_n\right), \tag{2.10}$$

where $G$ is an arbitrary function. The absence of flux (2.6) is caused by the dependence of the phase $\varphi_n$ (1.14) only on time. That is all solutions of equation (1.15) without magnetic field ( $\vec{A}_n=\vec{\theta}$ ) and with phase $\nabla_{\eta_n}\varphi_n=\vec{\theta}$ according to the Helmholtz decomposition (1.13)-(1.14) have zero flux, hence will lead to characteristic solutions (2.10). The specific form of the solution (2.9) is conditioned by the type of potential (2.3) and the time phase $\varphi_n=-E_\mu\tau_n/\hbar_n$ .

Integrating the solution (2.9) over $\int d\overset{(n-1)}{x}$ according to (1.11) will give the distribution function $f_{n-1}^\mu$ and equation (2.8) converts to the equation from the Vlasov chain

$$\frac{\partial f_{n-1}^\mu}{\partial t}+v\frac{\partial f_{n-1}^\mu}{\partial x}+\dot{v}\frac{\partial f_{n-1}^\mu}{\partial v}+...+\frac{\partial}{\partial \overset{(n-3)}{v}}\left[f_{n-1}^\mu\left\langle\overset{(n-2)}{v}\right\rangle^\mu\right]=0. \tag{2.11}$$

As an example let us find the function $f_{n-1}^\mu$ and the average flux $\left\langle v^{(n-2)}\right\rangle^\mu$ for the case $n=2$. Depending on the value of time $t$ the region of integration in expression (1.11) and (i.5) will be different. Figure 1 shows the area of integration for two fundamentally different times $t<a_2/\dot{a}_2$ (red color) and for $t>a_2/\dot{a}_2$ (blue color). In both cases the area of the phase domain (parallelogram) remains constant.

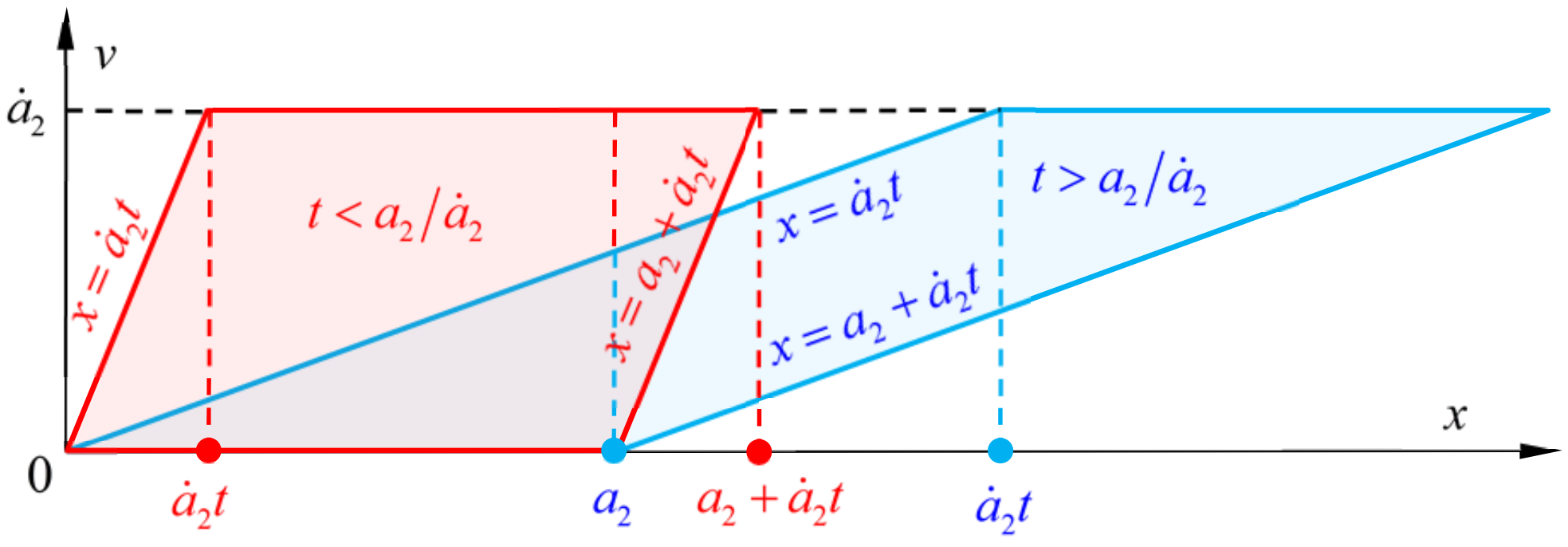

Fig. 1 Integration area

As a result, the expressions for the distributions $f_1^\mu$ and $\langle v\rangle^\mu$ take the following form (see Appendix)

$$f_1^\mu(x,t)=J_1^\mu(x,t),\ \langle v\rangle^\mu(x,t)=J_2^\mu(x,t), \tag{2.12}$$

where at $p=1,2$

$$J_p^\mu(x,t)=\begin{cases} I_p^\mu\left(0,\dfrac{x}{t},x,t\right),\ 0\le x<\dot a_2 t,\ t<\dfrac{a_2}{\dot a_2}\ \ or\ \ 0\le x<a_2,\ t\ge\dfrac{a_2}{\dot a_2},\\ I_p^\mu(0,\dot a_2,x,t),\ \dot a_2 t\le x<a_2,\ 0\le t<\dfrac{a_2}{\dot a_2},\\ I_p^\mu\left(\dfrac{x-a_2}{t},\dfrac{x}{t},x,t\right),\ a_2\le x<\dot a_2 t,\ t\ge\dfrac{a_2}{\dot a_2},\\ I_p^\mu\left(\dfrac{x-a_2}{t},\dot a_2,x,t\right),\ a_2\le x<\dot a_2 t+a_2,\ 0\le t<\dfrac{a_2}{\dot a_2}\ \ or\ \ \dot a_2 t\le x<\dot a_2 t+a_2,\ t\ge\dfrac{a_2}{\dot a_2},\end{cases} \tag{2.13}$$

$$a_2\dot a_2 I_1^\mu(v_1,v_2,x,t)\overset{\text{def}}{=}v_2-v_1-\frac{1}{\lambda_\mu t}\cos\{\lambda_\mu[2x-(v_1+v_2)t]\}\sin[\lambda_\mu(v_2-v_1)t], \tag{2.14}$$

$$\begin{aligned}2\lambda_\mu^2 a_2\dot a_2 t^2 f_1^\mu(x,t)I_2^\mu(v_1,v_2,x,t)=2\lambda_\mu^2 a_2\dot a_2 f_1^\mu(x,t)tx+\lambda_\mu^2 t(v_2-v_1)[(v_2+v_1)t-2x]+\\ +\lambda_\mu\bar v_1\sin[2\lambda_\mu(x-v_1t)]-\lambda_\mu\bar v_2\sin[2\lambda_\mu(x-v_2t)]+\\ +\sin\{\lambda_\mu[2x-(v_2+v_1)t]\}\sin[\lambda_\mu(v_1-v_2)t],\end{aligned}$$

where $\lambda_\mu=\pi\mu/a_2$. Figure 2 shows the evolution of the density $f_1^\mu(x,t)$ (2.12) for the state $\mu=5$. On the left side of Fig. 2 is an isometric projection of the distribution of $f_1^5(x,t)$ and on the right side is a top view (level lines). The blue color corresponds to the minimum values of the function and the red color corresponds to the maximum values of the function. In the white-colored area in Fig. 2 on the right are the function values $f_1^5>f_{fix}$, where $f_{fix}$ is the threshold value chosen for better image contrast.

At the initial moment of time the number of peaks in the distribution $f_1^5$ is equal to the state number $\mu=5$. Then they become four, then three, two and one. As a result the probability density becomes blurred. In Fig. 2 on the right we can see that the area of blurring grows according to the horizontal size of the parallelepiped in Fig. 1. Note that despite the simple form of the original distribution (2.6) after transformations (2.9) and (2.12) the distribution $f_1^\mu(x,t)$ has quite complex behavior.

In Fig. 3 shows the evolution of the velocity projection of the probability stream $\langle v\rangle^\mu(x,t)$ (2.12) for the state with the number $\mu=5$. The velocity projection is positive, i.e. the flow moves towards the right edge of the well. As a result the right edge of the wave has a very sharp front (red in Fig. 3) while the left edge is characterized by a delayed blur (blue in Fig. 3). Similar to the density behavior (see Fig. 2) the number of velocity peaks also decreases with time due to the blurring of the distribution. Note that although initially (2.6) the mean acceleration flux $\langle\dot v\rangle^\mu$ was zero the mean velocity flux $\langle v\rangle^\mu$ (2.12) has a complex inhomogeneous structure along the coordinate (see Fig. 3).

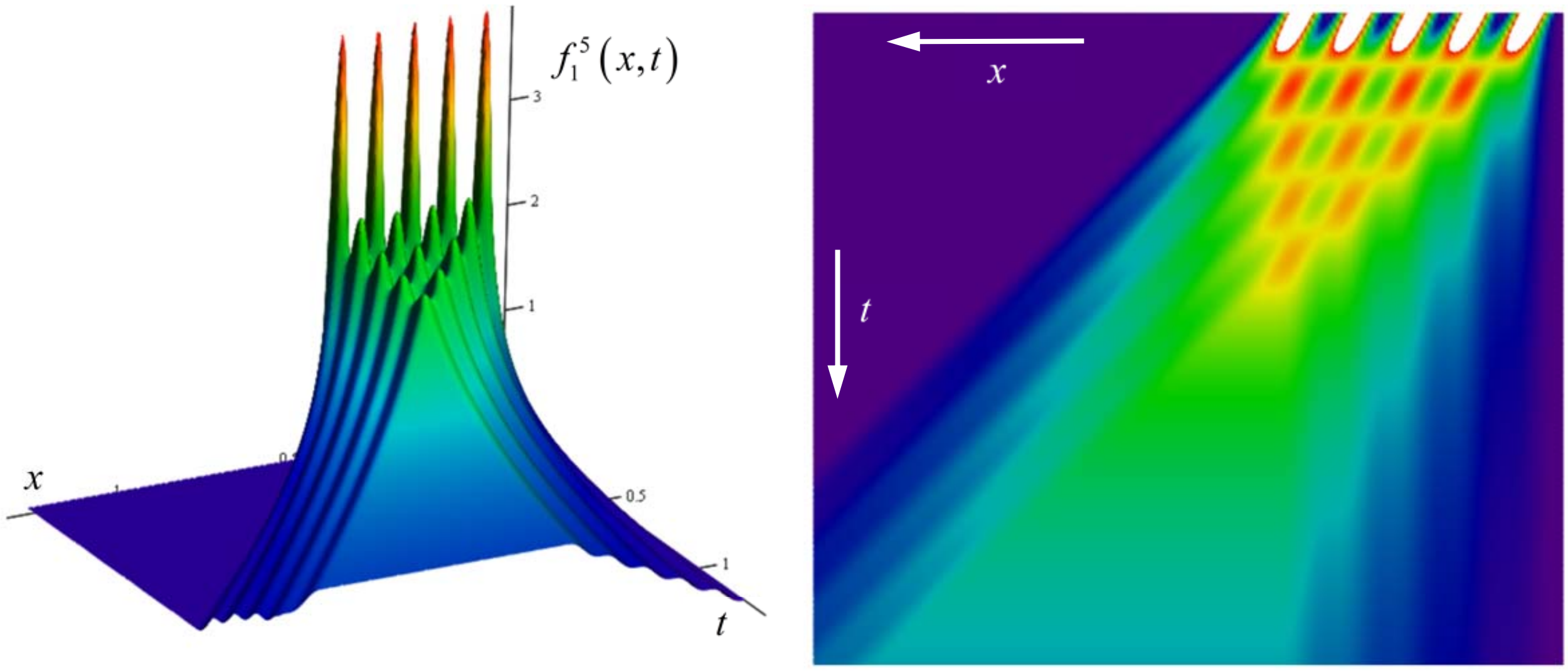

Fig. 2 Evolution of the distribution density $f_1^5(x,t)$

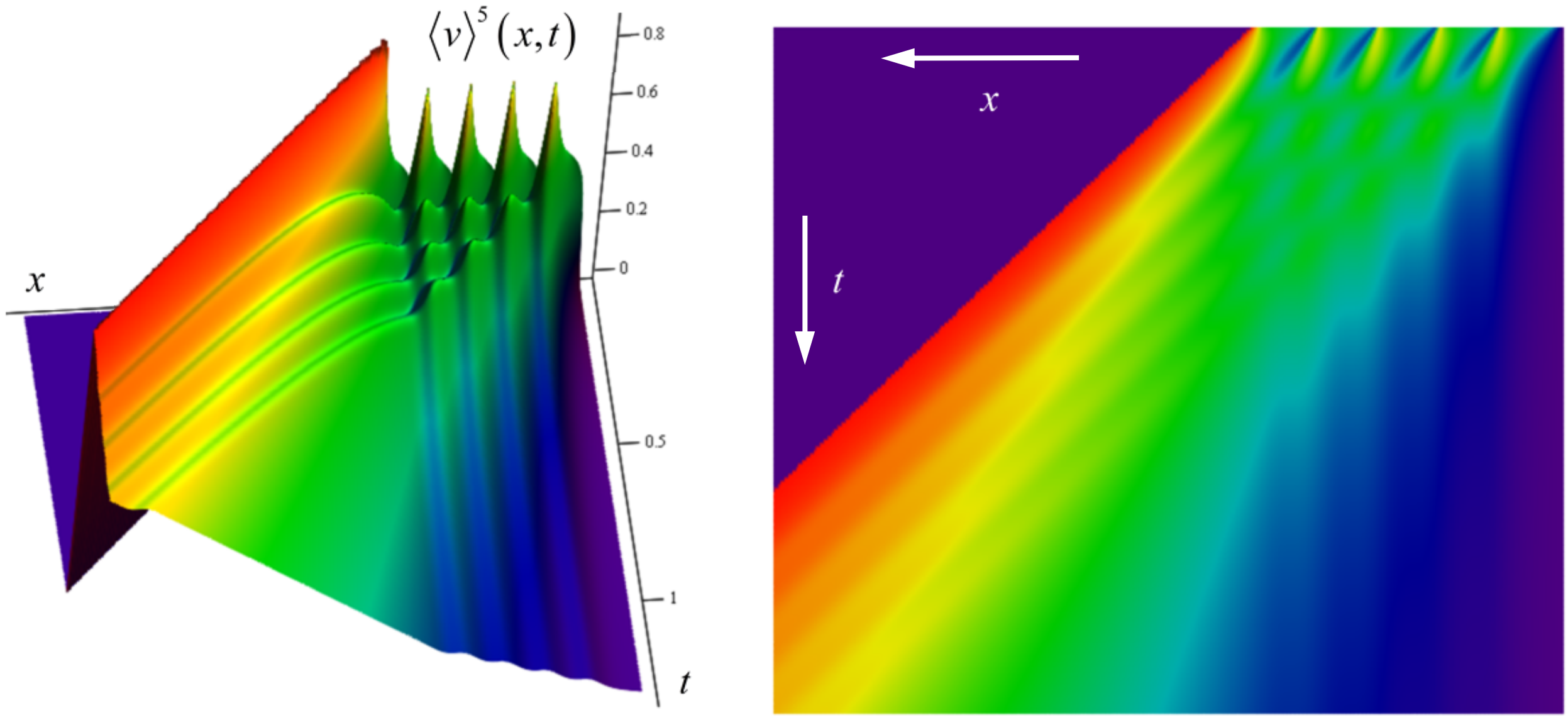

Fig. 3 Evolution of the probability flow rate $\langle v\rangle^5(x,t)$

Note that expression (2.5) is not the only solution of problem (2.4). In [32] a new solution of problem (2.4) was constructed which has a nonzero flux (1.13) and a time-dependent probability density function represented as the so-called «Dirac comb». The solution of problem (2.5) has the form

$$\Psi_n^{\mu,\beta}(\eta_n,\tau_n)=\frac{1}{\sqrt{N(\beta)}}\theta_1\left(\mu\frac{\eta_n}{a_n},-\mu^2\frac{2\pi\hbar_n}{ma_n^2}\tau_n+i\beta\right),\tag{2.15}$$

$$N(\beta)=a_n\sum_{k=-\infty}^{+\infty}e^{-\frac{\pi\beta}{2}(2k+1)^2},\qquad \int_0^{a_n}\left|\Psi_n^{\mu,\beta}(\eta_n,\tau_n)\right|^2 d\eta_n=1,$$

where the parameter $\beta>0$ and $\theta_1$ is the Jacobi theta-function

$$\theta_1(z,\tau)=\sum_{k=-\infty}^{+\infty} e^{i\frac{\pi\tau}{4}(2k+1)^2+i\frac{\pi}{2}(2z+1)(2k+1)},\ z\in\mathbb{C},\ \tau=\alpha+i\beta,\ \alpha,\beta\in\mathbb{R}. \tag{2.16}$$

The solution of (2.15) admits a representation in the form of

$$\Psi_n^{\mu,\beta}(\eta_n,\tau_n)=\frac{1}{\sqrt{N(\beta)}}\theta_1\left(\frac{\sqrt{2m\varepsilon_\mu}}{\hbar_n}\eta_n,-\frac{4\pi\varepsilon_\mu}{\hbar_n}\tau_n+i\beta\right),\qquad \varepsilon_\mu=\frac{\hbar_n^2\mu^2}{2ma_n^2},\quad E_\mu=\pi^2\varepsilon_\mu. \tag{2.17}$$

The distribution function $F_n^{\mu,\beta}$ corresponding to the solution (2.15) in contrast to (2.6) depends on time

$$F_n^{\mu,\beta}(\eta_n,\tau_n)=\left|\Psi_n^{\mu,\beta}\right|^2=\frac{1}{N(\beta)}\sum_{s,k=-\infty}^{+\infty} e^{-\frac{\pi\beta}{4}\left[(2k+1)^2+(2n+1)^2\right]}T_{|k-s|}\left(\cos\left[\vartheta_{s,k}^{\mu}(\eta_n,\tau_n)\right]\right), \tag{2.18}$$

$$\vartheta_{s,k}^{\mu}(\eta_n,\tau_n)\overset{\text{def}}{=}\pi\left(2\mu\frac{\eta_n}{a_n}+1\right)-\frac{\pi\tau_n}{\mathrm{T}_\mu}(s+k+1), \tag{2.19}$$

$$\mathrm{T}_\mu\overset{\text{def}}{=}\frac{\pi\hbar_n}{4E_\mu}=\frac{ma_n^2}{2\pi\hbar_n\mu^2}, \tag{2.20}$$

where $\mathrm{T}_\mu$ is the period of the function (2.17), and $T_{|k-s|}$ is the Chebyshev polynomials. Expression (2.19) defines the characteristics in the form of straight lines with slope angle $\theta$

$$\frac{\eta_n}{a_n}=\frac{s+k+1}{2\mu}\frac{\tau_n}{\mathrm{T}_\mu}+const\ \Rightarrow\ \mathrm{tg}\,\theta=\frac{s+k+1}{2\mu}, \tag{2.21}$$

along which the summands in the series (2.18) are constant.

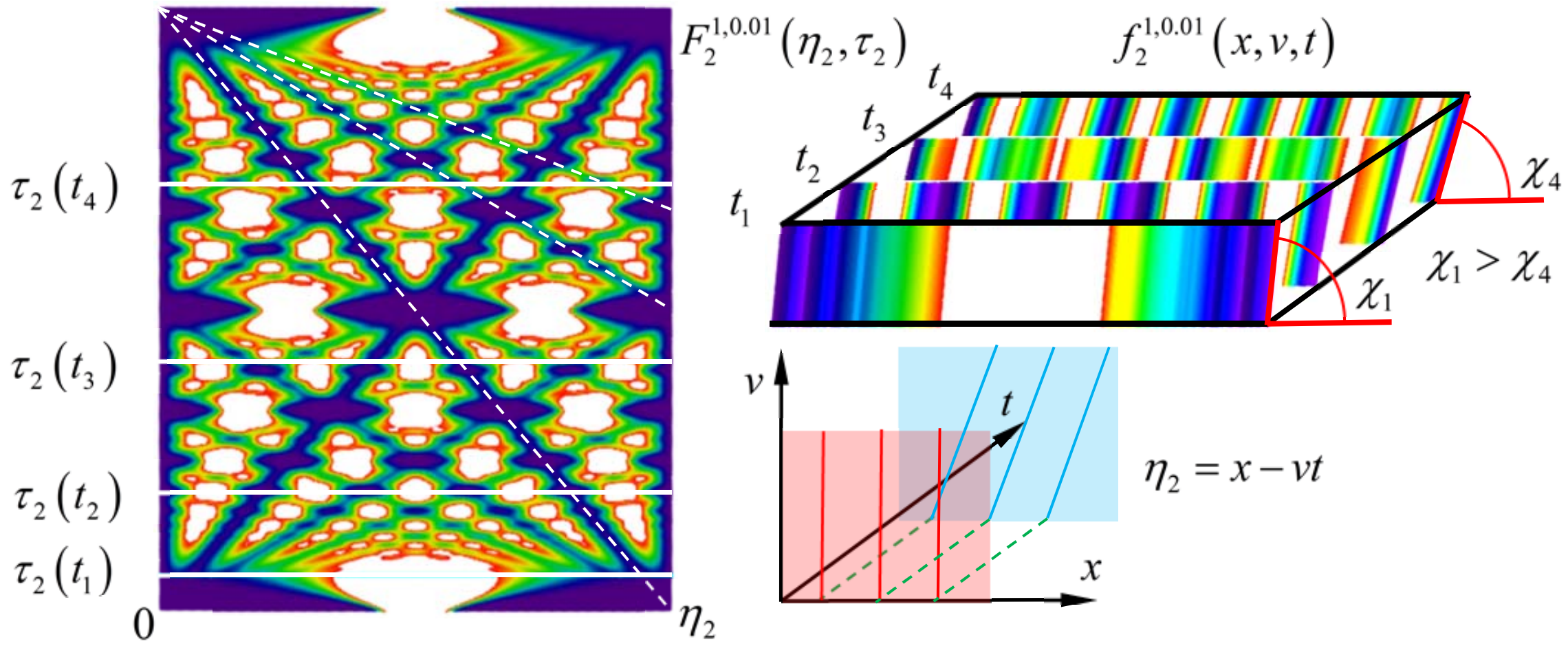

Fig. 4 Evolution of the characteristic solution in phase space

Fig. 4 (left) shows the distribution function view (2.18) of $F_2^{1,0.01}(\eta_2,\tau_2)$, i.e. $\mu=1$, $\beta=0.01$, $n=2$ and $a_2=0.5$. The blue color corresponds to the minimum values and the red color corresponds to the maximum values. Since the minimum and maximum values of the function $F_2^{1,0.01}$ are very different from each other for the contrast of the image of the level lines in Fig. 4 (left) is set to the maximum allowable value $F_{fit}$. As a result in the white regions in Fig. 4 (left) the function value $F_2^{1,0.01} > F_{fit}$. Horizontally in Fig. 4 (left) are plotted the coordinate values $0 < \eta_2 < a_2$ and vertically the time variable $0 < \tau_2 < \mathrm{T}_1$. The dotted white lines show the features (2.21) with different slope angles $\theta$.

The expression for the flux (1.18) unlike (2.6) is non-zero

$$\left\langle \overset{(n-1)}{u} \right\rangle^{\mu,\beta}(\eta_n,\tau_n) = \frac{a_n}{2\mu \mathrm{T}_\mu F_n^{\mu,\beta}(\eta_n,\tau_n) N(\beta)} \times \\ \times \sum_{s,k=-\infty}^{+\infty} e^{-\frac{\pi\beta}{4}\left[(2k+1)^2+(2s+1)^2\right]} (s+k+1)\cos\left[\vartheta_{s,k}^{\mu}(\eta_n,\tau_n)(k-s)\right]. \tag{2.22}$$

In Fig. 5 we show the flux distributions (2.22) $\langle \dot{u} \rangle^{1,0.01}(\eta_2,\tau_2)$ inside the potential well (2.3) at different time instants $\tau_2$ corresponding to the first half of the period $\mathrm{T}_1$. At the initial moments of time $0 < \tau_2 < \mathrm{T}_1/3$ in the right part of the well $0.25 < \eta_2 < 0.5$ the flow $\langle \dot{u} \rangle^{1,0.01}$ has a positive direction (from the center to the right edge of the well) and in the left part of the well $0 < \eta_2 < 0.25$ the flow direction $\langle \dot{u} \rangle^{1,0.01}$ is negative (from the center of the well to the left edge of the well). Closer to the middle of the period $\tau_2 = \mathrm{T}_1/2.5$, $\mathrm{T}_1/2.2$ flows in opposite directions reflected from the well walls appear. The dotted line in Fig. 5 shows the fluxes having regions with positive and negative directions. In the second part of the period $\mathrm{T}_1/2 < \tau_2 < \mathrm{T}_1$ the flux distributions $\langle \dot{u} \rangle^{1,0.01}$ in Fig. 5 will change to the opposite, i.e. to $-\langle \dot{u} \rangle^{1,0.01}$. Thus, after the period $\mathrm{T}_1$ the distributions $\langle \dot{u} \rangle^{1,0.01}$ and $F_2^{1,0.01}$ will repeat again.

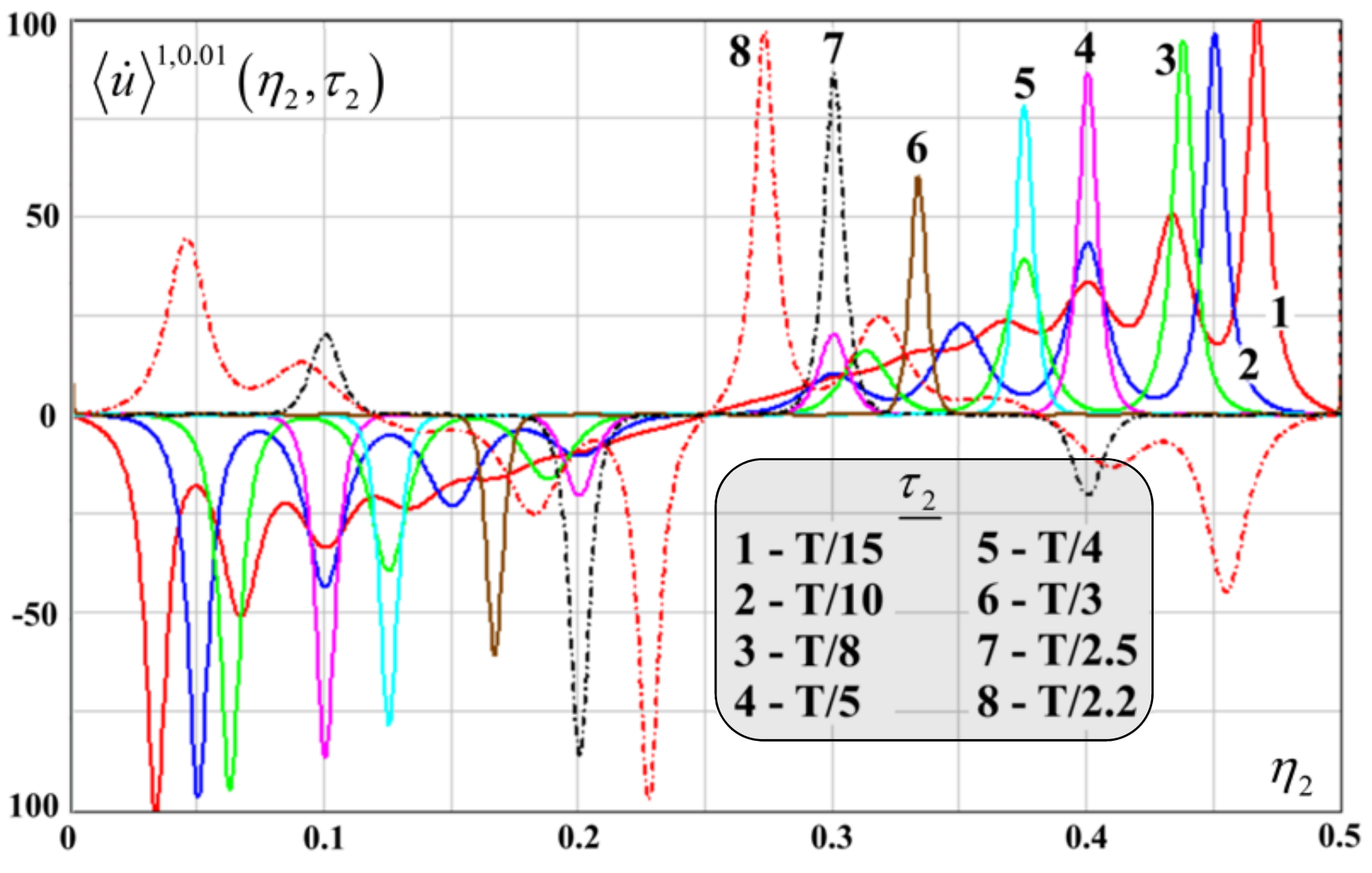

Fig. 5 Evolution of the acceleration flow

**Remark 3**

Let us consider the physical meaning of the parameter $\beta$. If we perform the limit transition at $\beta \to +\infty$ for the function $F_n^{\mu,\beta}(\eta_n,\tau_n)$ then we obtain the known distribution (2.6), i.e.

$$\lim_{\beta\to+\infty} F_n^{\mu,\beta}(\eta_n,\tau_n) = F_n^{\mu}(\eta_n). \tag{2.23}$$

The similar result is also true for the function $\bar{F}_n^{\mu,\beta}(\eta_n)$ averaged over the period $\mathrm{T}_\mu$

$$\bar{F}_n^{\mu,\beta}(\eta_n) \stackrel{\text{def}}{=} \frac{1}{\mathrm{T}_\mu}\int_0^{\mathrm{T}_\mu} F_n^{\mu,\beta}(\eta_n,\tau_n)\,d\tau_n = \frac{2}{N(\beta)}\sum_{k=-\infty}^{+\infty} e^{-\frac{\pi\beta}{2}(2k+1)^2}\sin^2\left[(2k+1)\frac{\pi\mu}{a_n}\eta_n\right],$$
$$\lim_{\beta\to+\infty}\bar{F}_n^{\mu,\beta}(\eta_n) = F_n^{\mu}(\mu), \tag{2.24}$$

where $\frac{\pi\mu}{a_n} = \frac{\sqrt{2mE_\mu}}{\hbar_n}$. The limit transition $\beta\to+\infty$ for the flow (2.22) leads to its vanishing

$$\lim_{\beta\to+\infty}\left\langle \overset{(n-1)}{u}\right\rangle^{\mu,\beta}(\eta_n,\tau_n) = 0. \tag{2.25}$$

In [32] it was shown at ($n=1$) that the parameter $\beta$ can be interpreted as the inverse temperature of the system. «Cooling» of the system ($\beta\to+\infty$) leads to damping of the probability fluxes (2.25) and to «freezing» of the probability densities (2.23) and (2.24). In this case the energy spectrum of the initial time-dependent system (2.18) asymptotically comes out to the known spectrum $E_\mu$ of the time-independent system (2.5).

Substitution of expressions (2.18) and (2.22) into equation (2.2) gives the correct identity. Consequently, we can construct the characteristic solution of the original equation (2.1)

$$f_n^{\mu,\beta}\left(x,\dot{x},\ddot{x},\ldots,\overset{(n-1)}{x},t\right) = F_n^{\mu,\beta}\left(\sum_{k=0}^{n-1}\overset{(k)}{x}\frac{(-1)^k t^k}{k!},\frac{(-1)^{n+1}t^n}{n!}\right). \tag{2.26}$$

The normalization condition for the function (2.26) is similar to (2.7) since the function $\Psi_n^{\mu,\beta}$ is normalized by (2.15). That is despite the explicit dependence on the time parameter $\tau_n$ of the density function $F_n^{\mu,\beta}$ its norm does not depend on $\tau_n$.

Fig. 4 (right) shows the distributions of $f_2^{1,0.01}(x,v,t_j)$ at equally spaced time moments $t_j$, $j=1\ldots4$. The time moments $t_j$ correspond to non-uniform steps in the time parameter $\tau_2(t_j)$ (1.3). At the initial time instant $t=0$ the phase region in the space $(x,v)$ is a rectangle with angle $\chi=\pi/2$ (see Fig. 4 right). The features $\eta_2 = const$ (1.3) are perpendicular to the axis $OX$. The time $t$ determines the angle of $\chi$ characteristics (see Fig. 4 right). Changing the slope angle $t_j$ of $\chi_1>\chi_4$ characteristics $\eta_2 = const$ over time causes the shape of the phase domain $(x,v)$ to change from a rectangle to a parallelogram (red and blue plane in Fig. 4 right).

**Remark 4**

The transition from $n$-th equation in the Vlasov chain (i.3) to $n-1$-th equation is made by integration over the corresponding phase subspace (i.5). In this transition it is implied [14, 15] that the distribution function $f_n$ has a sufficiently fast decay to zero at the outer boundary (infinity at $\mathbb{R}^3$). As a result by the Ostrogradsky-Gauss theorem the integral from the divergence of the current density reduces to zero flux at the outer boundary.

Note that the solutions (2.9) and (2.26) are zero not on the whole boundary of the phase region. However, the solution (2.9) has zero flux (2.6) which is characteristic of the vast majority of solutions to the Schrödinger equations because of the type of phase $\varphi = -Et/\hbar$. As a result the zero flux (2.6) leads to zero current density at the boundary of the region in the Ostrogradsky-Gauss theorem and $n$-th equation (2.8) naturally passes into $n-1$-th equation (2.11) from the Vlasov chain. In contrast the solution of (2.26) has a non-zero mean flow (2.22), which allows us to focus on solving the $n$-th equation forgetting about the transition to the $n-1$-th equation.

## Conclusion

The proposed characteristic method allows us to construct exact solutions for classical and quantum systems. Exact model solutions are used in the design and optimization of complex physical systems as a «zero» approximation. For example, the model problem with potential (2.3) is the simplest approximation when describing such objects as quantum dots [35-37]. The presence of a time-dependent solution (2.18), (2.22), (2.24) significantly complicates the behavior of the system compared to the solution (2.5), (2.9). Although even for the simplest system (2.9) the transition to the solution (2.12) leads to a significant nonlinear effect (see Figs. 2, 3).

Let's note that for quantum systems the solution of the second Vlasov equation in the form of the Wigner function [26-28] has a specificity concerning classical systems. The matter is that the Wigner function is a function of quasi-density of probabilities, i.e. it has regions with negative values. According to Hudson's theorem [38] and its extension to the 3D case [39], only the Gaussian wave function leads to a positive Wigner function in the entire phase domain. The characteristic solutions obtained in this paper are always positive due to the construction of $F_n = |\Psi_n|^2$ which simplifies their treatment for classical systems.

The solutions (2.22), (2.18) and their special case (2.6) satisfy the Hamilton-Jacobi equation (1.16) and the corresponding equation of motion (1.18). Solutions (2.24) with the field (2.20) where (1.4) is satisfied satisfy the conservation law (i.8). In the simplest case of zero flux (2.6) the equation of motion (1.18) takes the form $\nabla_{\eta_n} \mathrm{V}_n = \nabla_{\eta_n} U + \nabla_{\eta_n} \mathrm{Q}_n = 0$, where the quantum potential is $\mathrm{Q}_n = E_\mu$. In the presence of flux (2.22) the equation of motion (1.18) takes the form $\frac{d}{dt}\left\langle u^{(n-1)} \right\rangle^{\mu,\beta} = -\frac{1}{m}\frac{\partial \mathrm{Q}_n^{\mu,\beta}}{\partial \eta_n}$ where the flux exists under the quantum pressure force.

## Appendix

### *Proof of Theorem 1*

Taking into account the independence of kinematic quantities, the Vlasov equation (1.1) will take the following form

$$\frac{\partial f_n}{\partial t} + \vec{v} \cdot \nabla_r f_n + \dot{\vec{v}} \cdot \nabla_v f_n + \ldots + f_n \operatorname{div}_{\overset{(n-2)}{v}} \left\langle \overset{(n-1)}{\vec{v}} \right\rangle + \left\langle \overset{(n-1)}{\vec{v}} \right\rangle \cdot \nabla_{\overset{(n-2)}{v}} f_n = 0. \tag{A.1}$$

Let's calculate the derivatives included in equation (A.1), we obtain

$$\frac{\partial f_n}{\partial t}=\frac{\partial F_n}{\partial \tau_n}\frac{\partial \tau_n}{\partial t}+\frac{\partial F_n}{\partial \eta_n^{(x)}}\frac{\partial \eta_n^{(x)}}{\partial t}+\frac{\partial F_n}{\partial \eta_n^{(y)}}\frac{\partial \eta_n^{(y)}}{\partial t}+\frac{\partial F_n}{\partial \eta_n^{(z)}}\frac{\partial \eta_n^{(z)}}{\partial t}=$$
$$=(-1)^{n+1}\frac{t^{n-1}}{(n-1)!}\frac{\partial F_n}{\partial \tau_n}+\left[-\vec{v}+\dot{\vec{v}}t-...+(-1)^{n+1}\overset{(n-2)}{\vec{v}}\frac{t^{n-2}}{(n-2)!}\right]\cdot\nabla_{\eta}F, \tag{A.2}$$

$$\nabla_r f_n=\nabla_{\eta_n}F_n,\ \nabla_v f_n=-t\nabla_{\eta_n}F_n,\ \nabla_{\dot{v}} f_n=\frac{t^2}{2}\nabla_{\eta_n}F_n,...,\nabla_{\overset{(n-3)}{v}} f_n=(-1)^{n+1}\frac{t^{n-2}}{(n-2)!}\nabla_{\eta_n}F_n, \tag{A.3}$$
$$\nabla_{\overset{(n-2)}{v}} f_n=(-1)^{n+1}\frac{t^{n-1}}{(n-1)!}\nabla_{\eta_n}F_n,$$

$$\operatorname{div}_{\overset{(n-2)}{v}}\left\langle \overset{(n-1)}{\vec{v}}\right\rangle=\operatorname{div}_{\overset{(n-2)}{v}}\left\langle \overset{(n-1)}{\vec{u}}\right\rangle=\frac{\partial}{\partial \eta_n^{(x)}}\left\langle \overset{(n-1)(x)}{u}\right\rangle\frac{\partial \eta_n^{(x)}}{\partial \overset{(n-2)(x)}{v}}+\frac{\partial}{\partial \eta_n^{(y)}}\left\langle \overset{(n-1)(y)}{u}\right\rangle\frac{\partial \eta_n^{(y)}}{\partial \overset{(n-2)(y)}{v}}+$$
$$+\frac{\partial}{\partial \eta_n^{(z)}}\left\langle \overset{(n-1)(z)}{u}\right\rangle\frac{\partial \eta_n^{(z)}}{\partial \overset{(n-2)(z)}{v}}=(-1)^{n+1}\frac{t^{n-1}}{(n-1)!}\operatorname{div}_{\eta_n}\left\langle \overset{(n-1)}{\vec{u}}\right\rangle. \tag{A.4}$$

Substituting expressions (A.2)-(A.4) into equation (A.1), we obtain

$$(-1)^{n+1}\frac{t^{n-1}}{(n-1)!}\frac{\partial F_n}{\partial \tau_n}+\left[-\vec{v}+\dot{\vec{v}}t-...+(-1)^{n+1}\overset{(n-2)}{\vec{v}}\frac{t^{n-2}}{(n-2)!}\right]\cdot\nabla_{\eta}F+$$
$$+\vec{v}\cdot\nabla_{\eta_n}F_n-t\dot{\vec{v}}\cdot\nabla_{\eta_n}F_n+\frac{t^2}{2}\ddot{\vec{v}}\cdot\nabla_{\eta_n}F_n+...+(-1)^{n+1}\frac{t^{n-2}}{(n-2)!}\overset{(n-2)}{\vec{v}}\cdot\nabla_{\eta_n}F_n+$$
$$+F_n\operatorname{div}_{\overset{(n-2)}{v}}\left\langle \overset{(n-1)}{\vec{v}}\right\rangle+\left\langle \overset{(n-1)}{\vec{u}}\right\rangle\cdot\nabla_{\overset{(n-2)}{v}} f_n=0,$$
$$(-1)^{n+1}\frac{t^{n-1}}{(n-1)!}\frac{\partial F_n}{\partial \tau_n}+F_n(-1)^{n+1}\frac{t^{n-1}}{(n-1)!}\operatorname{div}_{\eta_n}\left\langle \overset{(n-1)}{\vec{u}}\right\rangle+(-1)^{n+1}\frac{t^{n-1}}{(n-1)!}\left\langle \overset{(n-1)}{\vec{u}}\right\rangle\cdot\nabla_{\eta_n}F_n=0,$$
$$\frac{\partial F_n}{\partial \tau_n}+f_n\operatorname{div}_{\eta_n}\left\langle \overset{(n-1)}{\vec{u}}\right\rangle+\left\langle \overset{(n-1)}{\vec{u}}\right\rangle\cdot\nabla_{\eta_n}F_n=0. \tag{A.5}$$

Theorem 1 is proved.

Let us directly integrate the density function $f_n$ over the kinematic quantity space $\int d^3\overset{(n-1)}{r}$, we obtain

$$f_{n-1}\left(\vec{\xi}^{1,...,n-1},t\right)=\int_{-\overset{(n-1)}{z_0}}^{\overset{(n-1)}{z_0}} d\overset{(n-1)}{z}\int_{-\overset{(n-1)}{y_0}}^{\overset{(n-1)}{y_0}} d\overset{(n-1)}{y}\int_{-\overset{(n-1)}{x_0}}^{\overset{(n-1)}{x_0}} \mathrm{F}_n\left(-\sum_{s=0}^{n-2}\tau_s\overset{(s)}{x}-\tau_{n-1}\overset{(n-1)}{x},\eta_n^{(y)},\eta_n^{(z)},\tau_n\right)d\overset{(n-1)}{x}=$$
$$=-\frac{1}{\tau_{n-1}}\int_{-\sum_{s=0}^{n-2}\tau_s\overset{(s)}{x}+\tau_{n-1}\overset{(n-1)}{x_0}}^{-\sum_{s=0}^{n-2}\tau_s\overset{(s)}{x}-\tau_{n-1}\overset{(n-1)}{x_0}} d\eta_n^{(x)}\int_{-\overset{(n-1)}{z_0}}^{\overset{(n-1)}{z_0}} d\overset{(n-1)}{z}\int_{-\overset{(n-1)}{y_0}}^{\overset{(n-1)}{y_0}} \mathrm{F}_n\left(\eta_n^{(x)},-\sum_{s=0}^{n-2}\tau_s\overset{(s)}{y}-\tau_{n-1}\overset{(n-1)}{y},\eta_n^{(z)},\tau_n\right)d\overset{(n-1)}{y}=$$

$$
=\frac{1}{\tau_{n-1}^{2}}\int\limits_{-\sum\limits_{s=0}^{n-2}\tau_s \overset{(s)}{y}+\tau_{n-1}\overset{(n-1)}{y_0}}^{-\sum\limits_{s=0}^{n-2}\tau_s \overset{(s)}{y}-\tau_{n-1}\overset{(n-1)}{y_0}} d\eta_n^{(y)} \int\limits_{\eta_{n,+}^{(x)}\left(\vec{\xi}^{1,\dots,n-1},t\right)}^{\eta_{n,-}^{(x)}\left(\vec{\xi}^{1,\dots,n-1},t\right)} d\eta_n^{(x)} \int\limits_{-\overset{(n-1)}{z_0}}^{\overset{(n-1)}{z_0}} \mathrm{F}_n\left(\eta_n^{(x)},\eta_n^{(y)},-\sum_{s=0}^{n-2}\tau_s \overset{(s)}{z}-\tau_{n-1}\overset{(n-1)}{z},\tau_n\right) d\overset{(n-1)}{z} =
$$

$$
=-\frac{1}{\tau_{n-1}^{3}}\int\limits_{\eta_{n,+}^{(y)}\left(\vec{\xi}^{1,\dots,n-1},t\right)}^{\eta_{n,-}^{(y)}\left(\vec{\xi}^{1,\dots,n-1},t\right)} d\eta_n^{(y)} \int\limits_{\eta_{n,+}^{(x)}\left(\vec{\xi}^{1,\dots,n-1},t\right)}^{\eta_{n,-}^{(x)}\left(\vec{\xi}^{1,\dots,n-1},t\right)} d\eta_n^{(x)} \int\limits_{-\sum\limits_{s=0}^{n-2}\tau_s \overset{(s)}{z}+\tau_{n-1}\overset{(n-1)}{z_0}}^{-\sum\limits_{s=0}^{n-2}\tau_s \overset{(s)}{z}-\tau_{n-1}\overset{(n-1)}{z_0}} \mathrm{F}_n\left(\eta_n^{(x)},\eta_n^{(y)},\eta_n^{(z)},\tau_n\right) d\eta_n^{(z)}. \qquad \text{(A.6)}
$$

Using (1.11) calculate $f_1^{\mu}(x,t)$, we obtain

$$
\begin{aligned}
f_1^{\mu}(x,t) &= \frac{1}{\dot{a}_2}\int\limits_{v_1}^{v_2} F_2^{\mu}(x-vt)\,dv = -\frac{1}{\dot{a}_2 a_2 t}\int\limits_{\bar{v}_1}^{\bar{v}_2}\left[1-\cos\left(2\lambda_\mu \bar{v}\right)\right]d\bar{v} = \\
&= \frac{1}{\dot{a}_2 a_2 t}\left\{\bar{v}_1-\bar{v}_2-\frac{1}{\lambda_\mu}\cos\left[\lambda_\mu\left(\bar{v}_2+\bar{v}_1\right)\right]\sin\left[\lambda_\mu\left(\bar{v}_1-\bar{v}_2\right)\right]\right\},\ \bar{v}_j = x-v_j t,\ j=1,2,
\end{aligned}
\qquad \text{(A.7)}
$$

where the integration limits $v_j$ are defined according to Fig. 3. For the expression of the average velocity (i.5) we similarly obtain

$$
f_1^{\mu}(x,t)\langle v\rangle^{\mu}(x,t) = \frac{1}{\dot{a}_2}\int\limits_{v_1}^{v_2} vF_2^{\mu}(x-vt)\,dv = -\frac{1}{a_2\dot{a}_2 t^2}\int\limits_{\bar{v}_1}^{\bar{v}_2}\left(x-\bar{v}\right)\left[1-\cos\left(2\lambda_\mu \bar{v}\right)\right]d\bar{v} =
$$

$$
= -\frac{x}{a_2\dot{a}_2 t^2}\int\limits_{\bar{v}_1}^{\bar{v}_2}\left[1-\cos\left(2\lambda_\mu \bar{v}\right)\right]d\bar{v} + \frac{1}{a_2\dot{a}_2 t^2}\int\limits_{\bar{v}_1}^{\bar{v}_2}\bar{v}\left[1-\cos\left(2\lambda_\mu \bar{v}\right)\right]d\bar{v} = \frac{x}{t}f_1^{\mu}(x,t) +
$$

$$
+\frac{\bar{v}_2^2-\bar{v}_1^2}{2a_2\dot{a}_2 t^2} - \frac{1}{2\lambda_\mu a_2\dot{a}_2 t^2}\left[\bar{v}_2\sin\left(2\lambda_\mu \bar{v}_2\right)-\bar{v}_1\sin\left(2\lambda_\mu \bar{v}_1\right)+\frac{1}{2\lambda_\mu}\cos\left(2\lambda_\mu \bar{v}_2\right)-\frac{1}{2\lambda_\mu}\cos\left(2\lambda_\mu \bar{v}_1\right)\right],
$$

$$
\begin{aligned}
\langle v\rangle^{\mu}(x,t) &= \frac{x}{t}+\frac{\bar{v}_2^2-\bar{v}_1^2}{2a_2\dot{a}_2 t^2 f_1^{\mu}(x,t)} - \frac{\bar{v}_2\sin\left(2\lambda_\mu \bar{v}_2\right)-\bar{v}_1\sin\left(2\lambda_\mu \bar{v}_1\right)}{2\lambda_\mu a_2\dot{a}_2 t^2 f_1^{\mu}(x,t)} + \\
&+\frac{\sin\left[\lambda_\mu\left(\bar{v}_2+\bar{v}_1\right)\right]\sin\left[\lambda_\mu\left(\bar{v}_2-\bar{v}_1\right)\right]}{2\lambda_\mu^2 a_2\dot{a}_2 t^2 f_1^{\mu}(x,t)},
\end{aligned}
\qquad \text{(A.8)}
$$

where (A.7) is taken into account.